\documentclass[prc,aps,eqsecnum,floatfix,nofootinbib]{revtex4-1}
\usepackage{bm}
\usepackage{xcolor}
\usepackage{graphicx}
\usepackage{amssymb,amsmath}
\allowdisplaybreaks[2]
\usepackage{multirow}
\usepackage{subfigure}

\newcommand{\Li}{{}^6{\rm Li}}
\newcommand{\Hes}{{}^6{\rm He}}
\newcommand{\xx}{\boldsymbol{\xi}}
\newcommand{\hxx}{\hat{\xi}}
\newcommand{\ph}{\varphi}
\newcommand{\br}{\boldsymbol{r}}

\newcommand{\al}{\alpha}

\newcommand{\phm}{\phantom{-}}

\begin{document}

%\preprint{}

\title{Comparative study of $\Hes$ $\beta$-decay based on different similarity-renormalization-group-evolved chiral interactions}


\author{A.\ Gnech$^{\rm a}$, L.E.\ Marcucci$^{\rm b,c}$, R. Schiavilla$^{\rm d,a}$, and M.\ Viviani$^{\rm c}$}
%\thanks{}
\affiliation{
$^{\rm a}$\mbox{Theory Center, Jefferson Lab, Newport News, Virginia 23606, USA}\\
$^{\rm b}$\mbox{Department of Physics, Universit\`a di Pisa, 56127 Pisa, Italy}\\ 
$^{\rm c}$\mbox{INFN-Pisa, 56127 Pisa, Italy}\\
$^{\rm d}$\mbox{Department of Physics, Old Dominion University, Norfolk, Virginia 23529, USA}
}


\date{\today}

\begin{abstract}
We report on a study of the Gamow-Teller matrix element contributing to $\Hes$ $\beta$-decay with
similarity renormalization group (SRG) versions of momentum- and configuration-space
two-nucleon interactions.  These interactions are derived from two different formulations of chiral effective
field theory ($\chi$EFT)---without and with the explicit inclusion of $\Delta$-isobars.
We consider evolution parameters $\Lambda_{\rm SRG}$ in the range
between 1.2 and 2.0 fm$^{-1}$ and, for the $\Delta$-less case, also
the unevolved (bare) interaction.  The axial current contains one- and
two-body terms, consistently derived at tree level (no loops) in the two distinct $\chi$EFT
formulations we have adopted here. The $\Hes$ and $\Li$
ground-state wave functions are obtained from hyperspherical-harmonics (HH)
solutions of the nuclear many-body problem.  In $A\,$=$\,6$ systems, the HH method
is limited at present to treat only two-body interactions and non-SRG evolved
currents.  Our results exhibit a significant dependence on $\Lambda_{\text{SRG}}$
of the contributions associated with two-body currents, suggesting that a
consistent SRG-evolution of these is needed in order to obtain reliable estimates.
We also show that the contributions from one-pion-exchange currents depend
strongly on the model (chiral) interactions and on the momentum- or configuration-space
cutoffs used to regularize them. These results might prove helpful in clarifying
the origin of the sign difference recently found in No-Core-Shell-Model and
Quantum Monte Carlo calculations of the $\Hes$ Gamow-Teller matrix element.
\end{abstract}

%\keywords{}

\maketitle
\section{Introduction}
 
Nuclear $\beta$ decays have become, in recent years, a research topic of intense interest.
A quantitative understanding of these decays is crucial for a number of experimental
endeavors, including the program of experiments planned at the Facility for Rare Isotope
Beams (FRIB) to measure weak-interaction rates in nuclei, and neutrinoless double-$\beta$
decay experiments aimed at establishing the Dirac or Majorana nature of the neutrino.
In this context, of particular relevance are Gamow-Teller matrix elements (GTMEs). 
Shell-model calculations have typically failed to reproduce the measured
values of these, unless an effective (one-body) Gamow-Teller (GT) operator is used with
a nucleon axial coupling constant $g_A$ that is quenched by about $ 20$--30\% relative to its free
value~\cite{Chou1993,Engel2017}.  The shell model also yields
rather uncertain estimates~\cite{Engel2017} for the nuclear matrix elements entering
neutrinoless double-$\beta$ decay rates, which are proportional
to $g_A^4$.  Therefore an understanding of the origin of $g_A$-quenching is important,
as is a reliable estimate of the contributions from many-body terms in the weak current.

There have been indications~\cite{Pastore2018,Gysbers2019,King2020}
that $g_A$ quenching might originate from lack of correlations in shell-model wave functions,
and possibly from two-body axial current contributions that tend to decrease the matrix element
calculated with the leading one-body GT operator~\cite{Gysbers2019}.  In this context, it
is interesting to note that the Gysbers {\it et al.}~study~\cite{Gysbers2019} consistently finds these two-body
contributions to generally have the opposite sign relative to the leading GT contributions in
nuclei with mass number $A>3$.  This is in contrast to the results of
Refs.~\cite{Pastore2018,King2020}, in which the sign of the one-
and two-body contributions is the same, at least in light nuclei with mass number $A \leq 10$,
the only ones accessible at this time to Green's function Monte Carlo (GFMC) methods.
The origin of this difference is yet to be clarified.  Of course, the comparison of
results obtained by different groups is difficult owing to the different models
adopted to describe nuclear interactions, and the different methods used
to solve the nuclear quantum many-body problem.  At this point in time, what can
be stated with confidence is that Gysbergs {\it et al.}~\cite{Gysbers2019} and the authors of Refs.~\cite{Pastore2018,King2020} only agree on the magnitude of the
two-body corrections: they are small in the $ A \leq10$ mass range.

In this work, in an attempt to understand the origin of this discrepancy, we present a
calculation of the GTME contributing to the $\beta$-decay of $\Hes$, within the
hyperspherical-harmonics (HH) method developed by the Pisa group~\cite{Kievsky2008,Marcucci2019},
and recently extended to deal with $A\,$=$\,6$ nuclei~\cite{Gnech2020}. The $\Hes$ and $\Li$
wave functions are obtained from a Hamiltonian including two-nucleon ($2N$) interactions only.
Three-nucleon ($3N$) interactions are neglected, since it is not yet possible to incorporate them in
HH calculations of $A\,$=$\,6$ nuclei (although some progress in this direction has been recently
made, see Ref.~\cite{Schiavilla2021}, by including the $3N$ contact interaction that enters pionless
effective field theory at leading order).

We adopt $2N$ interactions obtained in two different formulations of $\chi$EFT: one~\cite{Entem2017}
includes pions and nucleons as degrees of freedom, while the other~\cite{Piarulli2015,Piarulli2016}
also includes $\Delta$-isobars.  To each of these, we apply the similarity renormalization group
(SRG) unitary transformation~\cite{Bogner2007} in order to accelerate the convergence rate of
the HH expansion.  In reference to the nuclear axial currents, we use
the chiral models of Refs.~\cite{Baroni2016} and~\cite{Baroni2018}
in conjunction with the $\Delta$-less and $\Delta$-full interactions, respectively. 
These currents are treated without applying the proper SRG transformations.
Clearly, the absence of both $3N$ interactions and the proper SRG-evolution of
interactions and currents does not allow us to obtain a complete and fully consistent description
of the process.  Nevertheless, having an independent method that
can deal with different interactions could prove helpful in clarifying  the origin
of some of the tensions mentioned above.

The main goal of the present work is to understand the origin of
the difference in sign obtained for the two-body contributions to the GTME of $\Hes$
$\beta$-decay in the no-core shell model (NCSM) and GFMC calculations,
reported in Ref.~\cite{Gysbers2019} and Refs.~\cite{Pastore2018,King2020}, respectively.
Since use of the next-to-next-to-leading-order (N2LO450) interaction of Ref.~\cite{Entem2017}
allows us to achieve a satisfactory convergence in the $A\,$=$\,6$ HH calculation even without
implementing the SRG transformation, we are also in a position to assess the impact of the SRG
evolution itself on the GTME, at least as it relates to the $2N$ interaction.  However, we
should note that the $2N$ interaction adopted here
and in the study of Ref.~\cite{Gysbers2019} are not the same; specifically, the authors of that
work use the next-to-next-to-next-to-next-to-leading-order (N4LO500) rather than the N2LO450
model of Ref.~\cite{Entem2017}.  The former is of higher order (N4LO versus N2LO) in the power counting and
has a slightly larger cutoff (500 MeV) than the latter (450 MeV).

The other $2N$ interaction we and the authors of Ref.~\cite{King2020}
use in the GTME calculations is the NV2-Ia model of Ref.~\cite{Piarulli2016}. 
For this interaction, however, in order to reach convergence in the HH expansion,
we are forced to implement the SRG transformation.  We consider four different values for the evolution
parameter $\Lambda_{\text{SRG}}$, namely $\Lambda_{\text{SRG}}\,$=$\,1.2$, 1.5, 1.8, and 2.0
fm$^{-1}$.  This allows us to disentangle how two-body axial-current
contributions are affected by the input $2N$ interaction model (whether N2LO450 or NV-Ia)
and by the corresponding SRG-evolved versions of these models. 

The paper is organized as follows. In Secs.~\ref{sec:ax3} and~\ref{sec:HH}
we provide a concise review of, respectively, interactions and axial currents, and
the HH approach for $A\,$=$\,6$ nuclei.  We report our results in Sec.~\ref{sec:results},
and close in Sec.~\ref{sec:conclusions} with some
concluding remarks.  A number of more technical issues having to do with the 
convergence of the HH method for the $\Li$ and the $\Hes$ ground states are
relegated to Appendices~\ref{app:class} and~\ref{app:conv}. 

\section{Interactions and axial currents}
\label{sec:ax3}

In this work we use two different $2N$ chiral interactions.  
The first one is the next-to-next-to-leading-order (N2LO) model by Entem,
Machleidt and Nosyk~\cite{Entem2017}.
This interaction is derived from a $\chi$EFT including pions and nucleons
as degrees of
freedom.  It is regularized in momentum space (with a cutoff
$\Lambda\,$=$\,450$ MeV),
and is strongly non-local in configuration space.

The second interaction is the next-to-next-to-next-to-leading-order (N3LO)
model developed in Refs.~\cite{Piarulli2015,Piarulli2016}, which includes,
in addition to
pion and nucleon, $\Delta$-isobar degrees of freedom.  It is formulated in
configuration space and is regularized in this space with two
regulators, one ($R_S$) for the short-range components associated with
$2N$
contact terms, and the other ($R_L$) for the long-range ones induced by one-
and
two-pion exchange.  Various combinations of $R_S$ and $R_L$ regulators are
available, but in this work we have selected the model denoted as NV2-Ia with
$(R_S,R_L)\,$=$\,(0.8,\!1.2)$ fm.

Below, we will refer to these two interactions as the E and P models by the
initial of the first author on the relevant publications, respectively
Ref.~\cite{Entem2017}
and~\cite{Piarulli2015}.  Both models are evolved using the SRG unitary
transformation~\cite{Bogner2007},
in order to improve the convergence of the HH calculation.
This SRG evolution leads to
momentum-space interactions which are transformed back to coordinate space
by standard Fourier transforms. The matrix elements are then computed using the
procedure of Ref.~\cite{Gnech2020}.

Since one of our
goals is to understand
the effect of these SRG-evolved interactions on the GTME, we
consider four different values 
for the evolution parameter $\Lambda_{\rm SRG}$, namely,
$\Lambda_{\rm SRG}\,$=$\,1.2$, 1.5, 1.8, 2.0
fm$^{-1}$.  Furthermore,
it has been possible with the E interaction to obtain reasonable convergence
without implementing any SRG evolution (that is, with the ``bare''
interaction).  This
has allowed us to compare directly the bare and SRG calculated GTME,
and to assess the role of SRG evolution on this observable (see below).
However, we do not account for $3N$ interactions, since
SRG evolution for these is not yet available.

Accompanying each of these interactions is a set of N3LO axial currents
derived consistently in $\chi$EFT---the formulation that includes pions
and nucleons for the E model, and that with, in addition, $\Delta$ isobars for
the P model.   We provide below their configuration-space expressions in
the limit of
vanishing momentum transfer of interest here:
\begin{itemize}
\item  The leading-order (LO) term consists of the Gamow-Teller operator
\begin{equation}
  {\bf A}^{\rm LO}_{i,a}=-\frac{g_A}{2}\, \tau_{i,a}\,  {\bm \sigma}_i \ ,
  \label{eq:axlo}
\end{equation}
and scales, in a two-body system, as $Q^{-3}$ in the power counting---here, $Q$ denotes generically a low-momentum scale;
\item The N2LO terms (scaling as $Q^{-1}$) consist of a relativistic correction
  to the Gamow-Teller operator
\begin{equation}
\label{eq:axrc}
{\bf A}^{\rm N2LO}_{i,a}({\rm RC})=\frac{g_A}{4\, m^2}\,\tau_{i,a}
\, {\bf p}_i\times \left( {\bm \sigma}_i \times {\bf p}_i\right)\ ,
\end{equation}
and of a two-body operator induced by a $\Delta$-isobar intermediate state
(this only enters the calculations
based on the P interaction)
\begin{eqnarray}\label{eq:deltaj}
{\bf A}^{\rm N2LO}_{ij,a}(\Delta)&=&
- \left({\bm \tau}_i\times{\bm \tau}_j\right)_a \left[
I_1(r_{ij};\alpha_1^\Delta)\, {\bm \sigma}_i \times {\bm \sigma}_j
+I_2(r_{ij};\alpha_1^\Delta)\, {\bm \sigma}_i\times \hat{\bf r}_{ij}\,\, {\bm \sigma}_j\cdot \hat{\bf r}_{ij}\right] 
\nonumber\\
&&- \tau_{j,a}
\left[ I_1(r_{ij};\alpha_2^\Delta) \, {\bm \sigma}_j
+I_2(r_{ij};\alpha_2^\Delta) \,\hat{\bf r}_{ij}\,\,
 {\bm \sigma}_j\cdot \hat{\bf r}_{ij} \right]+(i \rightleftharpoons j)\ ;
\end{eqnarray}
\item The N3LO terms (scaling as $Q^{0}$) consist of a two-body operator
  associated with one-pion exchange (OPE)
\begin{eqnarray}\label{eq:opej}
{\bf A}^{\rm N3LO}_{ij,a}({\rm OPE})&=&
-\left({\bm \tau}_i\times{\bm \tau}_j\right)_a \left[
I_1(r_{ij};\alpha_1)\, {\bm \sigma}_i \times {\bm \sigma}_j
+I_2(r_{ij};\alpha_1)\, {\bm \sigma}_i\times \hat{\bf r}_{ij}\,\,{\bm \sigma}_j
\cdot \hat{\bf r}_{ij}\right] 
\\
&&-  \tau_{j,a}
\left[ I_1(r_{ij};\alpha_2) \, {\bm \sigma}_j
  +I_2(r_{ij};\alpha_2) \,\hat{\bf r}_{ij}\,\, {\bm \sigma}_j\cdot \hat{\bf r}_{ij}
  \right]
-\left({\bm \tau}_i\times{\bm \tau}_j\right)_a\,\frac{1}{2}
\left\{ {\bf p}_i \,\, , \,\, 
\widetilde{I}(r_{ij};\widetilde{\alpha})\, {\bm \sigma}_j\cdot\hat{\bf r}_{ij}
 \right\}+(i \rightleftharpoons j) \ , \nonumber
\end{eqnarray}
and of a two-body contact operator
\begin{equation}
{\bf A}^{\rm N3LO}_{ij,a}({\rm CT})= I_c(r_{ij};z_0)\,
\left({\bm \tau}_i\times{\bm \tau}_j\right)_a\,
\left({\bm \sigma}_i\times{\bm \sigma}_j\right)  \ .
\label{eq:axct}
\end{equation}
\end{itemize}
In Eqs.~(\ref{eq:axlo})--(\ref{eq:axct}),
${\bf p}_k\,$=$\, -i\, {\bm \nabla}_k$, ${\bm \sigma}_k$, and
${\bm \tau}_k$ are the momentum operator, and
Pauli spin and isospin operators of nucleon $k$, respectively,
$\left\{ \dots\, ,\, \dots\right\}$ denotes the anticommutator, and
${\bf r}_{ij}={\bf r}_i-{\bf r}_j$.
Charge-raising ($+$) or charge-lowering ($-$)
currents follow from ${\bf A}_{\pm} = {\bf A}_{x}\pm i\, {\bf A}_{y}$, where
the subscript specifies the isospin component.  In a many-body system,  the
one-body operators above are summed over the nucleons
$\sum_i {\bf A}_{i,a}$, while the two-body ones over the nucleon pairs
$\sum_{i< j} {\bf A}_{ij,a}$.

The correlation functions entering the OPE and CT currents and
corresponding to the E interaction are regularized by a momentum space cutoff
given by $C_\Lambda(k)={\rm e}^{-(k/\Lambda)^4}$.
They can be written as
\begin{eqnarray}
\label{eq:e13}
I^{\rm E}_1(r;\alpha_i^{\rm E})&=&-\frac{\alpha_i^{\rm E}}{\Lambda r}\int_0^\infty dx\,
\frac{x^3}{x^2+(m_\pi/\Lambda)^{\,2}} \, {\rm e}^{-x^4} \, j_1(x\Lambda r)\ , \\
\label{eq:e14}
I^{\rm E}_2(r;\alpha^{\rm E}_i)&=&\alpha_i^{\rm E} \int_0^\infty dx\,
\frac{x^4}{x^2+(m_\pi/\Lambda)^{\,2}} \, {\rm e}^{-x^4} \, j_2(x\Lambda r)   \ , \\
\widetilde{I}^{\rm E}(r; \widetilde{\alpha}^{\rm\, E} )&=&-
\widetilde{\alpha}^{\,{\rm E}} \int_0^\infty dx\,
\frac{x^3}{x^2+(m_\pi/\Lambda)^{\,2}} \, {\rm e}^{-x^4} \, j_1(x\Lambda r)  \ ,\\
I^E_c(r;z^{\rm E}_0)&=&z^{\rm E}_0\,\frac{\Lambda^3}{2\pi^2}\int_0^\infty dx\, x^2 \,
{\rm e}^{-x^4}\, j_0(x\Lambda r) \ ,
\label{eq:e2.9}
\end{eqnarray}
where the $j_n(z)$ are spherical Bessel functions,  the
$\alpha^{\rm E}_i$ and $\widetilde{\alpha}^{\rm \,E}$ denote the combinations of coupling constants defined as
\begin{equation}
  \alpha_1^{\rm E}=\frac{\Lambda^3}{4\,\pi^2} \,\frac{g_A}{{f}_\pi^{\,2}}
  \left({c}_4+\frac{1}{4\, {m}}\right) \ ,
\qquad
\alpha_2^{\rm E}=\frac{\Lambda^3}{2\pi^2} \,\frac{g_A\, c_3}{{f}_\pi^{\,2}}\ ,\qquad
\widetilde{\alpha}^{\rm\, E}=\frac{\Lambda^2}{8\,\pi^2} \,\frac{g_A}{{m}\,
  {f}_\pi^{\,2}}\ ,
\end{equation}
and $z^{\rm E}_0$ is the low-energy constant (LEC) that characterizes the
contact axial current (its determination
is discussed below); note that the $\alpha^{\rm E}_i$ are adimensional.
Here, $g_A$ is the nucleon axial coupling constant
($g_A\,$=$\, 1.2723$), $f_\pi$ is the pion-decay
constant ($f_\pi\,$=$\, 92.4$ MeV), and
$m_\pi$ and $m$ are the pion and nucleon masses, respectively.
The values of the LECs $c_3$ and $c_4$ depend on the interaction model
(either E or P) and
are listed in Table~\ref{tab:c3c4}.
\begin{table}[]
  \centering
  \begin{tabular}{ccc}
    \hline\hline
    & E-model & P-model \\ \hline
    $c_3$ &  $-3.61$       & $-0.79$ \\
    $c_4$ &  $\phm2.64$      & $\phm 1.33$ \\
    \hline\hline
  \end{tabular}
  \caption{Values of the LECs $c_3$ and $c_4$ associated with the
    E~\cite{Entem2017} and
    P~\cite{Piarulli2015} chiral interactions and used in the accompanying
    axial currents; they are in units of GeV$^{-1}$. These values are obtained
    from fits to $\pi N$ data without (E-model)  and with (P-model) the inclusion of $\Delta$-isobars.}
  \label{tab:c3c4}
\end{table}

The (regularized) correlation functions entering the $\Delta$, OPE, and CT
currents and
corresponding to the P interaction are
\begin{eqnarray}
\label{eq:e13p}
I^{\rm P}_1(r;\alpha_i^{\rm P})&=&-\alpha_i^{\rm P}\,(1+\mu)\, \frac{e^{-\mu}}{\mu^3}
\, C_{R_L}(r)  \ , \\
\label{eq:e14p}
I^{\rm P}_2(r;\alpha_i^{\rm P})&=& \alpha_i^{\rm P}\, (3+3\,\mu+\mu^2)\,
\frac{e^{-\mu}}{\mu^3} \, C_{R_L}(r)  \ , \\
\label{eq:e15p}
\widetilde{I}^{\rm \,P}(r; \widetilde{\alpha}^{\rm\, P} )&=& -
\widetilde{\alpha}^{\rm \,P}\,(1+\mu )\, \frac{e^{-\mu}}{\mu^2} \, C_{R_L}(r)  \ ,\\
\label{eq:e16p}
I^P_c(r;z^{\rm P}_0)&=&z^{\rm P}_0\,\frac{1}{\pi^{3/2}\, R_S^3} \, {\rm e}^{-(r/R_S)^2} \ ,
\label{eq:e2.14}
\end{eqnarray}
where $\mu\,$=$\, m_\pi r$, and
\begin{equation}
\label{eq:cfl}
C_{R_L}(r)  = 1 - \frac{1}{ (r/ R_L)^s\, 
{\rm e}^{(r-R_L)/a_L}+1} \ .
\end{equation}
Here, $a_L\,$=$\,R_L/2$, and the exponent
$s$ is taken as $s\,$=$\, 6$.  The $R_S$ and
$R_L$ values are
$(R_S,R_L)\,$=$\,(0.8,\!1.2)$ fm, consistently with the P model for the
nuclear interaction.
The correlation functions entering the $\Delta$-current of
Eq.~(\ref{eq:deltaj}) are the same as Eqs.~(\ref{eq:e13p}) and~(\ref{eq:e14p})
but with $\alpha_i^P \rightarrow \alpha_i^\Delta$.
The $\alpha_i^\Delta$ and $\alpha_i^{\rm P}$
combinations are defined as
\begin{eqnarray}
  \alpha_1^\Delta&=& \frac{g_A}{8\, \pi}\, \frac{m_\pi^3}{f_\pi^2}\,c_4^\Delta\ ,
  \qquad\qquad\qquad
\alpha_2^\Delta =\frac{g_A}{4\, \pi}\, \frac{m_\pi^3}{f_\pi^2}\,c_3^\Delta \ , \\
\alpha^{\rm P}_1&=& \frac{g_A}{8\, \pi}\, \frac{m_\pi^3}{f_\pi^2}
\left( c_4 +\frac{1}{4\, m}\right)\ , \qquad
\alpha^{\rm P}_2 =\frac{g_A}{4\, \pi}\, \frac{m_\pi^3}{f_\pi^2}\,c_3 \ , \qquad
\widetilde{\alpha}^{\rm \,P}=\frac{g_A}{16\, \pi}\, \frac{m_\pi^2}{m\, f_\pi^2}\ , 
\end{eqnarray}
with the LECs $c_3^\Delta$ and $c_4^\Delta$ given by
\begin{equation}
  c_3^\Delta=-\frac{h_A^2}{9\, m_{\Delta N}} \ , \qquad
  c_4^\Delta=\frac{h_A^2}{18\, m_{\Delta N}} \ ,
\end{equation}
where $h_A$ is the nucleon-to-$\Delta$ axial coupling constant
($h_A\,$=$\,2.74$) and $m_{\Delta N}$ is the $\Delta$-nucleon
mass difference ($m_{\Delta N}\,$=$\, 293.1$ MeV).

\begin{table}[bth]
    \centering
  \begin{tabular}{ccccc}
    \hline\hline
    Model & LO & N2LO+N3LO(OPE) & N3LO(CT) & $z_0$ [fm$^3$]\\
    \hline
    E-SRG1.2 & 0.9722 & $-$0.0121 & $-$0.0090 & 0.1104\\
    E-SRG1.5 & 0.9666 & $-$0.0095 & $-$0.0060 & 0.0610\\
    E-SRG1.8 & 0.9606 & $-$0.0053 & $-$0.0042 & 0.0392\\
    E-SRG2.0 & 0.9572 & $-$0.0021 & $-$0.0041 & 0.0370\\
    E-bare   & 0.9446 & $\phm$ 0.0086 & $-$0.0021 & 0.0193\\
    \hline
    P-SRG1.2 & 0.9728 & $\phm$0.0118 & $-$0.0335 & 0.4665\\
    P-SRG1.5 & 0.9679 & $\phm$0.0182 & $-$0.0348 & 0.3963\\
    P-SRG1.8 & 0.9620 & $\phm$0.0253 & $-$0.0363 & 0.3843\\
    P-SRG2.0 & 0.9584 & $\phm$0.0294 & $-$0.0368 & 0.3764\\    
    \hline\hline
  \end{tabular}

  \caption{Results for the  GTME of the tritium $\beta$-decay for the different
    components of the currents.  Columns labeled
    LO, and N3LO(CT) refer to the contributions given by the
    axial currents of Eqs.~(\ref{eq:axlo}), and (\ref{eq:axct}),
    respectively; the column labeled N2LO+N3LO(OPE) refers to the
    cumulative contribution of the axial currents of Eqs.~(\ref{eq:axrc}),
    (\ref{eq:opej}), and~(\ref{eq:deltaj}) in the P-model case.
    The N3LO(CT) cotribution   is determined by fitting the LEC $z_0$, used in
    Eqs.~(\ref{eq:e2.9}) and~(\ref{eq:e2.14}),
    to reproduce the observed  $\beta$-decay GTME of tritium,
    $\langle GT \rangle_{\rm exp}/\sqrt{3}=0.9511\pm0.0013$~\cite{Baroni2016}. 
  The calculated values of the LEC $z_0$ in fm$^{3}$ are reported as well in the last column of the table.}   
  \label{tab:z0}
\end{table} 
Finally, as per the determination of $z_0$, we note that this LEC is related to the
LEC $c_D$ that appears in the $3N$ contact interaction~\cite{Baroni2018}.  Since $3N$
interactions are altogether ignored in the present work, we fix directly $z_0$ so as
to reproduce the experimental value of the GTME in tritium $\beta$-decay,
$\langle GT \rangle_{\rm exp}/\sqrt{3}=0.9511\pm0.0013$~\cite{Baroni2016},
without concerning ourselves with the
connection between $z_0$ and $c_D$.  We do so for each of the SRG-evolved interactions
corresponding to the E and P models.
The numerical results for the GTME in tritium $\beta$-decay and the
fitted values of $z_0$ are reported in Table~\ref{tab:z0}.

\section{The Hyperspherical Harmonic method}\label{sec:HH}  

The $\Li$ and $\Hes$ wave functions have been expanded using the HH basis.
As reference set of Jacobi vectors for six equal-mass particles we use
\begin{equation}
  \begin{aligned}
    \xx_{1p}&=\sqrt{\frac{5}{3}}\left(\br_n-\frac{\br_m+\br_l+\br_k+\br_j+\br_i}{5}
    \right) \ ,\\
    \xx_{2p}&=\sqrt{\frac{8}{5}}\left(\br_m-\frac{\br_l+\br_k+\br_j+\br_i}{4}
    \right) \ ,\\
    \xx_{3p}&=\sqrt{\frac{3}{2}}\left(\br_l-\frac{\br_k+\br_j+\br_i}{3}
    \right) \ ,\\
    \xx_{4p}&=\sqrt{\frac{4}{3}}\left(\br_k-\frac{\br_j+\br_i}{2}
    \right) \ , \\
    \xx_{5p}&=\br_j-\br_i \ ,\label{eq:jacvec}
  \end{aligned}
\end{equation}
where $(i,j,k,l,m,n)$ indicates a generic permutation $p$  of the particles.
By convention, $p=1$ is chosen to correspond to $(1,2,3,4,5,6)$.
For a given choice of the Jacobi vectors, the hyperspherical
coordinates are given by the hyperradius $\rho$, which is independent on the
permutation $p$ of the particles and is defined as
\begin{equation}
  \rho=\sqrt{\sum_{i=1,N}\xi_{ip}^2}\,,\label{eq:rho}
\end{equation}
and by a set of variables, which in the Zernike and Brinkman
representation~\cite{Zernike1935,Fabre1983}, are the polar angles
$\hxx_{ip}=(\theta_{ip},\phi_{ip})$ of each Jacobi vector and the
four additional ``hyperspherical'' angles
$\ph_{jp}$, with $j=2,\dots,5$, defined as
\begin{equation}
  \cos \ph_{jp}=\frac{\xi_{jp}}{\sqrt{\xi_{1p}^2+\dots+\xi_{jp}^2}}\ .
  \label{eq:phiang}
\end{equation}
Here, $\xi_{jp}$ is the magnitude of the Jacobi vector $\xx_{jp}$.
The set of variables $\hxx_{1p},\dots,\hxx_{5p},\ph_{2p},\dots,\ph_{5p}$
is denoted hereafter as $\Omega_p$.
The expression of the generic $A=6$ HH function is
\begin{equation}\label{eq:hh6}
  \begin{aligned}
    {\cal Y}^{KLM}_{\mu}(\Omega_p)&=\big[((( Y_{\ell_1}(\hat \xi_{1p})
      Y_{\ell_2}(\hat \xi_{2p}))_{L_2} Y_{\ell_3}(\hat \xi_{3p}))_{L_3}
%     \\
%    &\times
      Y_{\ell_4}(\hat \xi_{4p}))_{L_4} Y_{\ell_5}(\hat \xi_{5p})
      \big]_{LM}\\
    &\times{\cal P}^{\ell_1,\ell_2,\ell_3,\ell_4,\ell_5}
    _{n_2,n_3,n_4,n_5}(\ph_{2p},\ph_{3p},\ph_{4p},\ph_{5p})\,,
  \end{aligned}
\end{equation}
where
\begin{equation}\label{eq:hh6b}
  \begin{aligned}
    &{\cal P}^{\ell_1,\ell_2,\ell_3.\ell_4,\ell_5}
    _{n_2,n_3,n_4,n_5}(\ph_{2p},\ph_{3p},\ph_{4p},\ph_{5p})\\
    &={\cal N}_{n_2}^{\ell_2,\nu_2}(\cos\ph_{2p})^{\ell_{2}}
    (\sin\ph_{2p})^{\ell_1}P_{n_2}^{\ell_1+1/2,\ell_2+1/2}(\cos 2\ph_{2p})
    \\
    &\times{\cal N}_{n_3}^{\ell_3,\nu_3}(\cos\ph_{3p})^{\ell_3}
    (\sin\ph_{3p})^{K_2}P_{n_3}^{\nu_2,\ell_3+1/2}(\cos 2\ph_{3p})\\
    &\times{\cal N}_{n_4}^{\ell_4,\nu_4}(\cos\ph_{4p})^{\ell_4}
    (\sin\ph_{4p})^{K_3}P_{n_4}^{\nu_3,\ell_4+1/2}(\cos 2\ph_{4p})\\
    &\times{\cal N}_{n_5}^{\ell_5,\nu_5}(\cos\ph_{5p})^{\ell_5}
    (\sin\ph_{5p})^{K_4}P_{n_5}^{\nu_4,\ell_5+1/2}(\cos 2\ph_{5p})\ ,
  \end{aligned}
\end{equation}
and $P^{a,b}_n$ are Jacobi polynomials.
The coefficients ${\cal N}_{n_j}^{\ell_j,\nu_j}$ are normalization factors
given explicitly by
\begin{equation}\label{eq:nhh}
  {\cal N}^{\ell_j,\nu_j}_{n_j}=
  \biggl[{2\nu_j\Gamma (\nu_j-n_j)n_j!\over
      \Gamma (\nu_j-n_j-\ell_j-{1\over 2})
      \Gamma (n_j+\ell_j+{3\over 2})}\biggr]^{1/2}\,,
\end{equation}
and we have defined
\begin{equation}
  K_j=\ell_j+2n_j+K_{j-1}\,,\qquad
  \nu_j=K_j+\frac{3}{2}j-1\,,
\end{equation}
with $K_1=\ell_1$ and $K_5=K$. The integer index $\mu$ labels the
set of hyperangular quantum numbers, namely
\begin{equation}\label{eq:qn3}
  \mu\equiv\{\ell_1,\ell_2,\ell_3,\ell_4,\ell_5,L_2,L_3,
  L_4,n_2,n_3,n_4,n_5\}\ .
\end{equation}
The wave function is constructed to have a well-defined  total
angular momentum $J$ and third component $J_z$, parity $\pi$ and isospin $T$
(in the following, we ignore the small admixtures between isospin states induced
by isospin-symmetry-breaking interactions).
Therefore, a complete basis of antisymmetrical hyperangular-spin-isospin
states is constructed as follows
\begin{equation}\label{eq:hhst0}
  \Psi^{KLSTJ\pi}_\al=\sum_{p=1}^{360}\Phi^{KLSTJ\pi}_\al(i,j,k,l,m,n)\ ,
\end{equation}
where the sum is over the 360 even permutations $p$ of the particles and
\begin{equation}
  \begin{aligned}
    &\Phi^{KLSTJ\pi}_\al(i,j,k,l,m,n)=
    \big \{{\cal Y}^{KLM}_{\mu}(\Omega_{p})
         [[[s_is_j]_{S_2} s_k]_{S_3}\\
    &\quad\times[[s_ls_m]_{S_4} s_n]_{S_5}]_S\big\}_{JJ_z}
    [[[t_it_j]_{T_2} t_k]_{T_3}
      [[t_lt_m]_{T_4} t_n]_{T_5}]_{TT_z}\,.
    \label{eq:hhst}
  \end{aligned}
\end{equation}
The functions ${\cal Y}^{KLM}_{\mu}(\Omega_{p})$ are the HH functions defined in
Eq.~(\ref{eq:hh6}), and $s_i$ $(t_i)$ denotes the spin (isospin) state
of nucleon $i$. Note that the coupling scheme of 
these spin and isospin states does not follow that of the hyperangular part.
This particular choice
simplifies the calculation of the interaction matrix elements.
The index $\al$ labels the possible sets of hyperangular, spin and
isospin quantum numbers compatible with the given values of
$K$, $L$, $S$, $T$, $J$, and $\pi$, namely
\begin{equation}\label{eq:alpha}
  \begin{aligned}
    \alpha\equiv\{&\ell_1,\ell_2,\ell_3,\ell_4,\ell_5,
    L_2,L_3,L_4,n_2,n_3,n_4,n_5,\\
    &S_2,S_3,S_4,S_5,T_2,T_3,T_4,T_5\}\,.
  \end{aligned}
\end{equation}
The parity of the state is defined by
$\pi=(-1)^{\ell_1+\ell_2+\ell_3+\ell_4+\ell_5}$; of course, we include in the
basis only those states having the parity of the nuclear
state under consideration.
By exploiting the sum over the permutation, the antisymmetry
on the wave function is imposed by the condition
\begin{equation}
  \ell_5+S_2+T_2=\text{odd}\,.
\end{equation}
This method generates linearly dependent HH states. However,
in the basis we only include independent states, obtained by 
calculating the norm matrix elements and by
implementing the Gram-Schmidt orthogonalization procedure
(the technique is described in Ref.~\cite{Gnech2020}). This drastically reduces
the number of states used in the expansion.

The final form of the six-nucleons bound state wave function can be written as
\begin{equation}\label{eq:wf}
  \Psi_6^{TJ\pi}=\sum_l\sum_{KLS,\alpha}c^{KLST}_{l,\alpha}
  f_l(\rho)\Psi^{KLSTJ\pi}_\al\,,
\end{equation}
where the sum is over the linearly independent antisymmetric
states $\alpha$, and
$c^{KLST}_{l,\alpha}$ are variational coefficients to be
determined.
The hyperradial functions $f_l(\rho)$ are chosen to be
\begin{equation}
  f_l(\rho)=\gamma^{15/2} \sqrt{\frac{l!}{(l+14)!}}\,\,\, 
  L^{(14)}_l(\gamma\rho)\,\,e^{-\gamma \rho/2} \ ,
  \label{eq:fllag}
\end{equation}
where $L^{(14)}_l(\gamma\rho)$ are Laguerre polynomials~\cite{AbramowitzStegun},
and $\gamma$ is a non-linear variational
parameter that is introduced so as to improve the convergence on $l$.
A typical range for $\gamma$ is 3.5--5.5 fm$^{-1}$ while the sum over $l$ is typically carried up to $l=20$.
The expansion coefficients $c^{KLST}_{l,\alpha}$ are determined by using
the Rayleigh-Ritz variational principle.  The resulting eigenvalue problem
is solved with the procedure of Ref.~\cite{Cullum1981}.

Even though the number of states is much reduced,
a brute force approach, in which the complete basis of independent states
up to a maximum $K$ is included, is not yet possible. For this reason, we
select subsets of basis states, separating them
in classes of convergence.  Within each class, we analyze the
convergence pattern in order to obtain a reliable extrapolation for the
binding energy.  A fairly detailed discussion of these classes for  $\Li$
is given in Ref.~\cite{Gnech2020}.  It is summarized here in Appendix~\ref{app:class}
along with a discussion of the classes of convergence for $\Hes$.
In the appendix, we also discuss the extrapolation
procedure, and provide tables exhibiting the convergence pattern,
within each class and for each nucleus, corresponding to the different interaction models.

\section{results}\label{sec:results}
The extrapolated binding energies for the $\Li$ and $\Hes$ ground states corresponding
to the E and P models are listed in Table~\ref{tab:be}. We stress again
that $3N$ interactions as well as many-body interactions
induced by the SRG transformation are not accounted for.  Nevertheless, the results
obtained with the SRG-evolved versions of the E and P models happen to be quite
close to the experimental values.
\begin{table}[bth]
  \centering
  \begin{tabular}{rcccc}
    \hline
    \hline
    & \multicolumn{2}{c}{$\Li$} & \multicolumn{2}{c}{$\Hes$}\\
    & E-model & P-model & E-model & P-model \\
    \cline{2-3}\cline{4-5}
    SRG1.2 & 32.19(1) & 32.40(1) & 28.96(1) & 29.10(1)\\
    SRG1.5 & 33.47(2) & 33.88(2) & 30.31(1) & 30.61(1)\\
    SRG1.8 & 33.33(5) & 33.85(8) & 30.25(3) & 30.64(3)\\
    SRG2.0 & 32.94(7) & 33.43(8) & 29.89(4) & 30.22(5)\\
    bare   & 30.33(20)&          & 27.51(23) &\\
    \hline
    Exp. & \multicolumn{2}{c}{31.99} & \multicolumn{2}{c}{29.27}\\
    \hline
    \hline
  \end{tabular}
  \caption{\label{tab:be} Extrapolated values for the $\Li$ and $\Hes$
    binding energies obtained with the SRG-evolved versions of the
    E and P interactions, corresponding to $\Lambda_{\rm SRG}\,$=$\,1.2$,
    1.5, 1.8, and 2.0 fm$^{-1}$, and without SRG
    evolution for the E interaction; in parentheses, are extrapolation errors
    (see appendices for a discussion of how these are estimated). For comparison, we also report the
    experimental binding energies
    from Ref.~\cite{Tilley2002}.}
\end{table}

We define the reduced GTME as
\begin{equation}
      {\text{RME} (K_L,K_H)}=\frac{\sqrt{2J_f+1}}{g_A}\frac{\langle \psi_{J_f,M}(K_L)|A_{+}^z |\psi_{J_i,M}(K_H)\rangle}{\langle J_i M,1 0|J_fM\rangle}\,,
\end{equation}
where $A_{+}^z$ is the $z$-component (at vanishing momentum transfer) of the
total charge-raising axial current given in Sec.~\ref{sec:ax3}, and $\langle J_i M, 10|J_f M\rangle$ is a
Clebsch-Gordan coefficient; note that the $\Hes$ and $\Li$ ground states have $J_i^{\pi_i}\,$=$\,0^+$ and
 $J_f^{\pi_f}\,$=$\,1^+$, respectively. This matrix element depends explicitly on the maximum value
of $K$ used in the HH expansion of the $\Hes$ ($K_H$) and $\Li$ ($K_L$)
wave functions. Its evaluation is carried out by Monte Carlo
integration with $\sim 30000$ configurations, which yields a statistical error
of the order of $\sim1\%$ on the individual components beyond LO of the axial current,
except for the $A^{z,{\rm N3LO}}_+({\rm OPE})$ component
because of accidental cancellations (see below).

\begin{figure}
  \centering
  \includegraphics[scale=0.5]{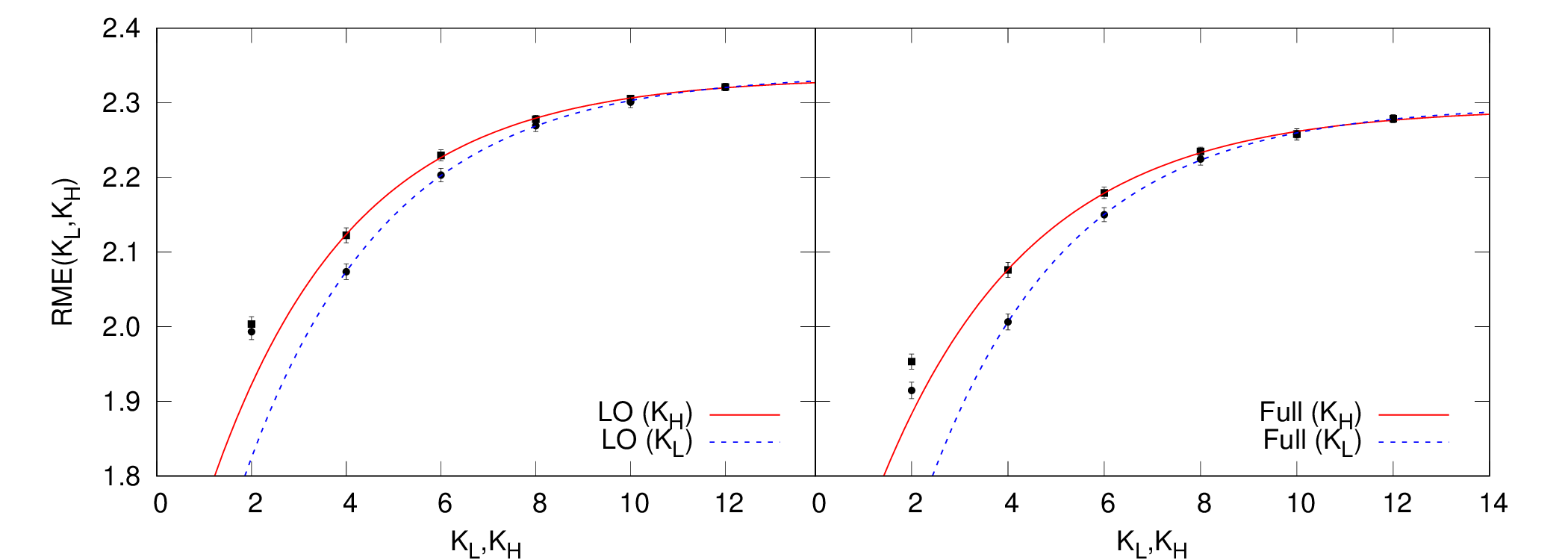}
  \caption{RME values computed as function of the maximum $K$ used in the
    expansion of the $\Li$ ($K_L$, circles) and $\Hes$ ($K_H$, squares)
    wave functions for the SRG2.0 version of the E interaction. The left (right) panel
    corresponds to results obtained with the LO (up to N3LO) axial current. The solid red (dashed blue) line
    is a fit to the calculated RMEs as function of $K_L$ ($K_H$); see text for further explanations.
    All remaining interactions exhibit a similar pattern of convergence.}
  \label{fig:conv}
\end{figure}

We study separately the convergence of the RME with respect to $K_H$
and $K_L$, since the states included in the HH expansions of the $\Hes$ and $\Li$
wave functions are different. We proceed as follows. We fix $K_L$ ($K_H$) to the
maximum value used in the present work---namely, $K_L\,$=$\,12$ ($K_H\,$=$\,12$)---and
then compute the matrix element by increasing the value of $K_H$ ($K_L$).  The LO
RME exhibits an exponential behavior with respect to both $K_H$
and $K_L$, as shown in the left panel of Fig.~\ref{fig:conv}.  We fit our results with a
function of the form $\text{RME}(K)\,$=$\,\text{RME}(\infty)+A\exp{(-bK)}$ for $K_{L,H}\ge4$, where
the parameter $\text{RME}(\infty)$ is the extrapolated value corresponding to
$K_H\longrightarrow\infty$ and $K_L\longrightarrow\infty$. The fits are indicated by the solid and dashed lines.
The two extrapolated values are then mediated with the weighted average in
order to obtain the final result. The same exponential behavior of the RME is
observed when all axial-current contributions up to N3LO are included, see
solid lines in the right panel of Fig.~\ref{fig:conv}.  It is worthwhile noting, though,
that this behavior is essentially driven by the LO term, since higher-order terms
only provide small corrections to the RME, see below.

\begin{table}[bth]
  \centering
  \begin{tabular}{lcccccc}
    \hline
    \hline
    \multicolumn{7}{c}{E-model}\\
    \hline
    & LO & N2LO(RC) & N3LO(OPE) & N3LO(CT) & N3LO(CT)/$z^{\rm E}_0$ [fm$^{-3}$] & Full \\
    SRG1.2 & 2.345(2) & $-$0.019 & $-$0.038(1) & $-$0.018 & $-$0.162(1) & 2.271(3) \\
    SRG1.5 & 2.342(3) & $-$0.021 & $-$0.029(1) & $-$0.011 & $-$0.185(1) & 2.281(2) \\
    SRG1.8 & 2.327(3) & $-$0.022 & $-$0.019(1) & $-$0.008 & $-$0.198(1) & 2.281(4) \\
    SRG2.0 & 2.338(3) & $-$0.022 & $-$0.013(1) & $-$0.008 & $-$0.202(1) & 2.297(2) \\
    bare   & 2.321(9) & $-$0.023 & $\phantom{-}$0.002(1)& $-$0.004 & $-$0.211(1) & 2.303(11) \\
    \hline
    \multicolumn{7}{c}{P-model}\\
    \hline
    & LO & N2LO(RC+$\Delta$) & N3LO(OPE) & N3LO(CT) & N3LO(CT)/$z^{\rm P}_0$ [fm$^{-3}$] & Full \\
    SRG1.2 & 2.354(1) & $-$0.033(1) & $\phantom{-}$0.011 & $-$0.066 & $-$0.143(1) & 2.265(2) \\
    SRG1.5 & 2.331(4) & $-$0.030(1) & $\phantom{-}$0.016 & $-$0.066 & $-$0.166(1) & 2.251(3) \\
    SRG1.8 & 2.329(5) & $-$0.023(1) & $\phantom{-}$0.020 & $-$0.068 & $-$0.177(1) & 2.257(4) \\
    SRG2.0 & 2.322(6) & $-$0.019(1) & $\phantom{-}$0.022 & $-$0.070 & $-$0.185(1) & 2.260(11) \\
    \hline
    NV2-Ia + 3b(VMC)~\cite{King2020} & 2.200 &$\phantom{-}$0.022 &$\phantom{-}$0.039 & $-$0.005 &$-$0.009 & 2.256 \\
    NV2-Ia + 3b(GFMC)~\cite{King2020} & 2.130 & & & &  & 2.201 \\
    Exp.~\cite{Knecht12}&  & & & & & 2.1609(40) \\
    \hline
    \hline
  \end{tabular}
  \caption{Extrapolated RMEs in $\Hes$ $\beta$-decay. The
    results are obtained by using SRG-evolved versions of the E and P models
    with $\Lambda_{\text{SRG}}=1.2,1.5,1.8,2.0$ fm$^{-1}$;
    $3N$ interactions are not included. Columns labeled
    LO, N2LO(RC), N3LO(OPE), and N3LO(CT) refer to the contributions given by the
    axial currents of Eqs.~(\ref{eq:axlo}), (\ref{eq:axrc}), (\ref{eq:opej}), and (\ref{eq:axct}),
    respectively; the column labeled N2LO(RC+$\Delta$) refers to the
    cumulative contribution of the axial currents of Eqs.~(\ref{eq:axrc}) and~(\ref{eq:deltaj}).
    Note that the N3LO(CT) results have been divided out by the $z_0$ values listed in
    Table~\ref{tab:z0}. The errors, when shown, are associated with the extrapolation;
    when they are not explicitly indicated, they are below the precision reported in
    the table. For a qualitative comparison, we also list the results of
    Ref.~\cite{King2020} obtained with the P model (including $3N$ interactions),
    and the experimental value from Ref.~\cite{Knecht12}.}
  \label{tab:RME}
\end{table}
Considering separately the contributions beyond LO, we observe that they
do not present any particular convergence pattern. However, the calculations
for $8\leq K_L,K_H\leq12$ are compatible within twice the statistical
error bars of the Monte Carlo integration.  Therefore, we consider as our best
estimate the weighted average between the values obtained in the range
$8\leq K_L,K_H\leq12$. We note that the convergence pattern of these contributions is
independent of the interaction model (either E or P) and the value of $\Lambda_{\text{SRG}}$.
The extrapolated values of the RME for each individual component of the
current as well as for the full current are reported in Table~\ref{tab:RME}.
We find that two-body currents give a overall correction of
opposite sign to the LO contribution,  of the order of $\sim 3/4\%$,
in line with the results of Refs.~\cite{Vaintraub2009,Gysbers2019}.
However, a closer inspection of the table suggests a more complex situation.

From the first column of Table~\ref{tab:RME}, the LO contribution seems to
have a weak dependence on $\Lambda_{\text{SRG}}$:  the
larger is $\Lambda_{\text{SRG}}$, the smaller is the resulting LO
contribution. This same sensitivity is also shown in Fig.~8 of
Ref.~\cite{Gysbers2019} and, as demonstrated by the authors of that paper,
it is removed by
including the SRG-induced two-body operators corresponding to the LO current. By
comparing the results for the bare E model with its SRG evolved versions, the
difference is of the order of $\sim0.5\%$ and we would have expected a similar difference
also in the case of the P model, had we been able to use the bare interaction.
However, the results reported in Ref.~\cite{King2020} show that at least a correction
of the order of $5\%$ is needed. This extra quenching of the LO contribution
comes from $3N$ interaction effects~\cite{WiringaPC}.
  
The N2LO(RC) contribution, which only consists of relativistic corrections
to the LO Gamow-Teller operator, appears to be independent of the
SRG-evolution parameter for both the E and P models.  By contrast, 
the N2LO($\Delta$) contribution strongly depends on $\Lambda_{\text{SRG}}$,
and is responsible for generating the pattern shown in Table~\ref{tab:RME}.
It starts off negative for $\Lambda_{\rm SRG}\,$=$\,1.2$ fm$^{-1}$,
increases monotonically, and becomes positive for $\Lambda_{\rm{SRG}}\,$=$\,2.0$ fm$^{-1}$. 
In Ref.~\cite{King2020} (with the bare P interaction) this contribution is found to
be positive and larger than the negative N2LO(RC) contribution,
resulting in an overall positive value for the sum.  Here, the
situation is reversed, and even at $\Lambda_{\rm SRG}\,$=$\,2.0$ fm$^{-1}$
the sum of the N2LO(RC) and N2LO($\Delta$) contributions remains negative.
%We conjecture that this might be due to
%the absence of $3N$ interactions in our calculation which can influence the
%relativistic corrections that have a short-range behavior.
Such a difference is clearly due to SRG-evolution effects.

The N3LO(OPE) contribution also depends strongly on
$\Lambda_{\text{SRG}}$, see Table~\ref{tab:RME}.  For the E model,
it starts off negative at low $\Lambda_{\text{SRG}}$, and increases monotonically
as $\Lambda_{\text{SRG}}$ increases, becoming positive in the limit
$\Lambda_{\text{SRG}}\longrightarrow \infty$, corresponding to the bare
interaction.  This is a clear indication that a proper SRG evolution of the
N3LO(OPE) current---as well as the N2LO($\Delta$) current, discussed above---is needed
to obtain reliable estimates. Such a program has been partially carried out in Ref.~\cite{Gysbers2019}.
However, to best of our knowledge, three body induced axial currents generated by the SRG evolution have not been included.
The results obtained with the P model show the same behavior as
function of $\Lambda_{\text{SRG}}$. However, in this case the N3LO(OPE) contribution
is positive for all $\Lambda_{\rm SRG}$ used, and the calculations seem to go
in the direction of Ref.~\cite{King2020} when $\Lambda_{\text{SRG}}$ increases.
However, we should point out that,
because of cancellations between the terms proportional to $c_3$ and $c_4$,
the overall N3LO(OPE) contribution is rather sensitive to the actual values of these LECs,
in particular their ratio $c_3/c_4$.
Lastly, for this contribution we do not expect significant effects from $3N$ interactions,
since the latter do not affect appreciably the short-range behavior of two-nucleon
densities.  These densities, and the resulting change of sign between the
E and P N3LO(OPE) contributions, are studied in the next section.  

The N3LO(CT) contributions are found to be negative for both interaction models.
When divided out by the LEC $z_0$---column labeled N3LO(CT)/$z_0$---they are
almost identical between the E and P models.  The results reported in Table~\ref{tab:RME}
exhibit a significant dependence on the SRG-evolution parameter. 
It is interesting to note, however, how these results, when they are multiplied by the
fitted values of $z_0$ from Table~\ref{tab:z0}, become essentially
independent of $\Lambda_{\text{SRG}}$ for the P model.  By contrast,
in the case of the E model the results remain $\Lambda_{\text{SRG}}$-dependent,
albeit the trend is inverted (rather than decreasing, they increase as $\Lambda_{\rm SRG}$
increases).  It seems that $z_0$ can absorb, at least
partially, the effect of the SRG evolution of the currents.
By comparing our results for the P model with those of
Ref.~\cite{King2020}, there is almost one order of magnitude of difference.
%{\color{red} We should understand the origin of this difference, I don't think $3N$ interactions are the culprit.}
  
\subsection{Two-body transition densities}
\begin{figure}
  \centering
  \includegraphics[scale=0.5]{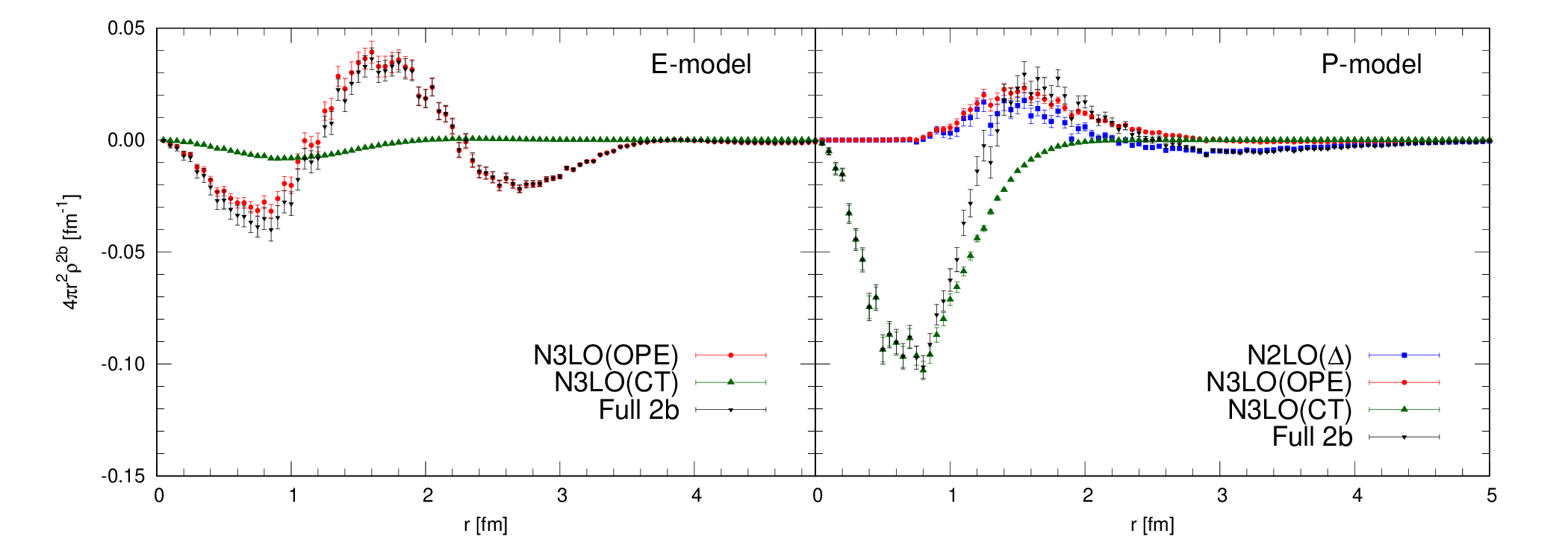}
  \caption{Two-body densities defined in Eq.~(\ref{eq:tbd}) for the SRG-evolved
  versions of the E and P interactions corresponding to 
    $\Lambda_{\text{SRG}}=2.0$ fm$^{-1}$. Similar results
    are obtained for all other $\Lambda_{\text{SRG}}$ considered in this work,
     except for the N2LO($\Delta$) contribution
for the P-model where for $\Lambda_{\text{SRG}}=1.2,1.5,1.8$ fm$^{-1}$ the
two-body transition densities result of opposite sign.}
  \label{fig:tbd0}
\end{figure}

In order to understand the differences between the results obtained with the
two different chiral interactions, we compute the two-body transition density,
which we define as~\cite{King2020,Schiavilla1998}
\begin{equation}\label{eq:tbd}
  \text{RME(2b)}=4\pi\int_0^\infty\, dr\,r^2\,\rho^{2b}(r)\,,
\end{equation}
where $r$ is the distance between two nucleons and 2b stands for N2LO($\Delta$)
(only for the P model), N3LO(OPE), and N3LO(CT).  In
Fig.~\ref{fig:tbd0} we report the two-body densities computed using
$\Lambda_{\text{SRG}}=2.0$ fm$^{-1}$ for the E and P models. Their shape
is independent on the $\Lambda_{\text{SRG}}$ value, except for the N2LO($\Delta$) contribution
for the P-model where for $\Lambda_{\text{SRG}}=1.2,1.5,1.8$ fm$^{-1}$ the
two-body transition densities result of opposite sign.

\begin{figure}[bth]
  \centering
  \includegraphics[scale=0.7]{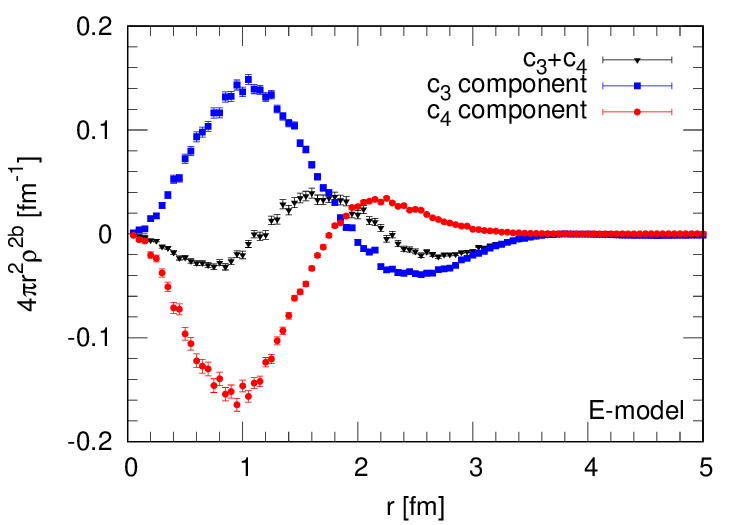}
  \includegraphics[scale=0.7]{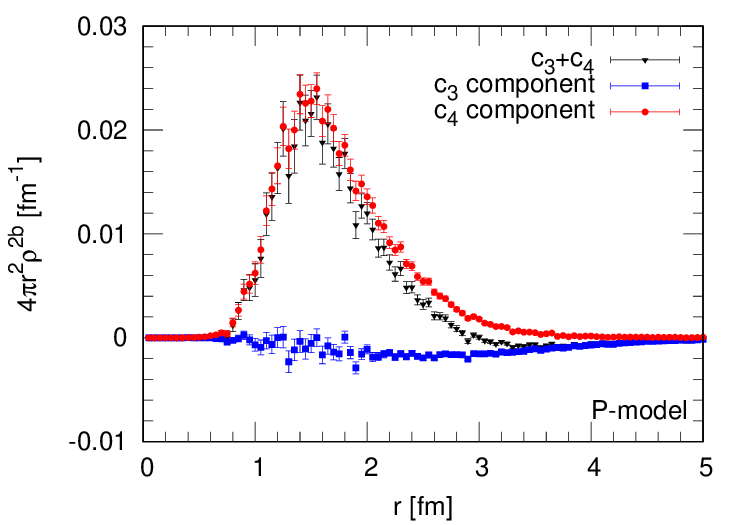}
  \caption{Two-body densities for the N3LO(OPE) contribution (solid points) computed with
  the SRG-evolved versions of the E and P interactions with
  $\Lambda_{\text{SRG}}\,$=$\,2.0$ fm$^{-1}$.  The
    blue (red) points indicate the density corresponding to the component proportional to $c_3$ ($c_4$) only.}
  \label{fig:tbd1}
\end{figure}

Inspection of the two panels in Fig.~\ref{fig:tbd0} indicates that 
the N3LO(OPE) densities corresponding to the E and P models
are rather different.  As a matter of fact, the
shape of these densities is determined by the cancellation between
the two components of the current proportional to the LECs
$c_3$ and $c_4$ through $\alpha_1$ and $\alpha_2$ in
Eq.~(\ref{eq:opej}).  In Fig.~\ref{fig:tbd1} we plot the separated contributions
for the two interactions. In the E model there is a double lobe structure for
both the $c_3$ and $c_4$
components.  This, and the fact that the maxima of the second lobes do not
coincide, generate a three-lobe structure with two of the lobes negative
and one positive. In the P model, the $c_3$ and $c_4$ components have both just
one lobe, which generates a single lobe in the total contribution (see Fig.~\ref{fig:tbd1}).
This is qualitatively consistent with the results reported in
Ref.~\cite{King2020}.

\begin{figure}
  \centering
  \includegraphics[scale=0.7]{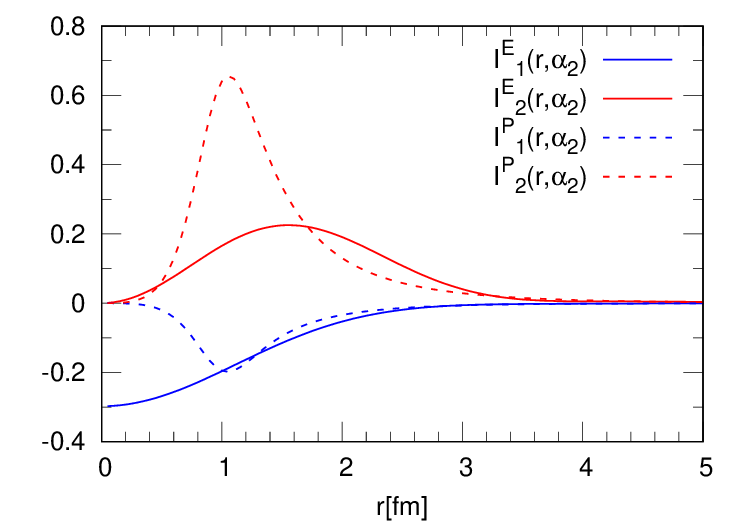}
  \caption{Correlation functions in the N3LO(OPE) current for the
    E model from Eqs.~(\ref{eq:e13})--(\ref{eq:e14}) (full lines), and the
    P model from Eqs.~(\ref{eq:e13p})--(\ref{eq:e14p}) (dashed lines).
    In the figure we show the $I_1$ and $I_2$ functions proportional to $c_3$ only, but
    with $c_3\,$ set to 1 in units of GeV$^{-1}$.}
  \label{fig:Icorr}
\end{figure}

The difference in the N3LO(OPE) densities of the E and P models originates from that in the
corresponding correlation functions entering the current, see Eqs.~(\ref{eq:e13})--(\ref{eq:e14})
and Eqs.~(\ref{eq:e13p})--(\ref{eq:e14p}).   We plot those proportional to $c_3$ (with $c_3\,$=$\,1$ in
units of GeV$^{-1}$
to make the comparison meaningful) in Fig.~\ref{fig:Icorr}.  In the region $r\lesssim 3$ fm,
their shapes are affected by the choice of regulator.  This also produces the sign inversion
between the E- and P-model $c_3$ (and $c_4$) contributions, shown in Fig.~\ref{fig:tbd1}.
%Indeed, the integrals of the two correlation functions are very different among the two interaction
%models, changing the relative weight for the $c_3$ and $c_4$ components of the OPE current.
%As it can be deduce  from Fig.~\ref{fig:Icorr}, the integrals
%of $I_1^E(r,\alpha_i)$ and $I_2^E(r,\alpha_i)$ are very similar, generating
%similar contribution for the $c_3$ and $c_4$ components. This is not the case
%for $I_1^P(r,\alpha_i)$ and $I_2^P(r,\alpha_i)$, where the second is much
%larger than the first one, favoring the $c_4$ component of the OPE current respect to the $c_3$ one.
%
% 
%N2LO($\Delta$) contribution for the P-model, which results to be identical to the OPE one, since the operator
%structure and the correlation functions are the same [see Eq.~(\ref{eq:deltaj})]. 
  
For the N3LO(CT) contribution, the main difference between the two interactions
is the presence of a second tiny lobe at $r\approx 2$ fm in the E model,
and the fact that the maximum is shifted towards larger $r$-values (around 1 fm)
compared to that in the P model. Also in this case, the origin of the differences among
the two-body densities comes from the different behavior of the correlation functions
given in Eqs.~(\ref{eq:e2.9}) and ~(\ref{eq:e2.14}). The
results obtained with the P model with $\Lambda_{\text SRG}=2.0$ for the N2LO($\Delta$), N3LO(OPE), and
N3LO(CT) densities are in qualitative agreement with those of Ref.~\cite{King2020}.
  
\section{Conclusions}\label{sec:conclusions}

In this work, we have reported on a study of the $\Hes$ GTME, using different chiral two-nucleon
interactions, the N2LO450~\cite{Entem2017} and NV2-Ia~\cite{Piarulli2016} models.
Both models have been evolved via SRG unitary transformations corresponding
to parameters $\Lambda_{\rm SRG}$ between 1.2 and 2.0 fm$^{-1}$.  We have neglected
$3N$ and SRG-induced many-nucleon interaction effects, as well as SRG-induced
many-body terms in the nuclear axial current.

The results are summarized in Table~\ref{tab:RME}.  We find for both
models that all axial-current terms beyond LO yield a cumulative contribution,
which (in magnitude) amounts to a 3\% correction of the LO Gamow-Teller contribution.
We also find this cumulative contribution to have the opposite sign of the LO one,
in agreement with the results of Refs.~\cite{Gysbers2019,Vaintraub2009}.
The contributions of two-body currents, in particular of N3LO(OPE), while small,
depend strongly on the parameter $\Lambda_{\rm SRG}$, suggesting that
a consistent evolution of these currents (together with the one-body current)
may be necessary in order to obtain reliable predictions.
The same conclusion
can also be drawn by considering the results for the tritium $\beta$-decay in
Table~\ref{tab:z0}.
  
We have been unable to reproduce the sign of the beyond-LO contributions
obtained in Ref.~\cite{King2020} with the bare NV2-Ia interaction.   This can
be traced back to differences in the contributions associated with the N2LO(RC) and
N3LO(CT) currents.  The origin of these differences is unclear.  We conjecture they might
be due to the absence, in the present HH calculation, of the multi-nucleon terms
induced by the SRG transformation in the interactions and currents.  By contrast, there is
qualitative agreement in the shape of the two-body transition
densities calculated here and in Ref.~\cite{King2020}. 
  
We have shown that the N3LO(OPE) contribution is opposite in sign for the
SRG-evolved N2LO450 and NV2-Ia interactions.  The corresponding
transition densities in Fig.~\ref{fig:tbd0} have different shapes, reflecting
the different behavior of the correlation functions entering the N3LO(OPE)
current, see Fig.~\ref{fig:Icorr}.  This behavior follows in turn from the different
choice of short-range regulators we have adopted in the N2LO450 and
NV2-Ia calculations.   We note in closing that the
sign difference in the N3LO(OPE) contribution obtained in
Refs.~\cite{Gysbers2019} and~\cite{King2020} may have a similar origin.

\section*{Acknowledgements}
We thank G.B.~King, S.~Pastore, and R.B.~Wiringa for a useful email exchange on the effect of
$3N$ interactions on $\beta$ decay matrix elements. We also thank M.~Piarulli for her help during
the implementation of the SRG version of the Norfolk potential.
This research is supported by the U.S. Department of Energy, Office
of Nuclear Science, under contracts DE-AC05-06OR23177 (A.G. and R.S.).
The calculations were made possible by grants of computing time
from the National Energy Research Supercomputer Center (NERSC).

\appendix
\section{Classes of convergence}\label{app:class}

In this Appendix we define the classes of convergence in which
we separate the HH states.  This definition is based
on a couple of criteria. The first one is that, as
the value $\ell_{\text{sum}}=\ell_1+\ell_2+\ell_3+\ell_4+\ell_5$
increases, so does the centrifugal barrier, which keeps nucleons
apart from each other, thus reducing the effect of correlations
induced by the nuclear interactions.  The second criterion accounts for the fact that
the $2N$ interaction favors two-body correlations
and so the HH states with non-zero quantum numbers
for the couple $(i,j)$ are privileged. These states
can be easily selected by imposing $\ell_i=0$ with $i=1,2,3,4$.
Furthermore, the HH states can also be classified on the basis of their
$LST$ quantum numbers (or partial waves). Indeed, in the $\Li$ and $\Hes$
nuclei the most important partial waves are the $S$ and $D$ waves, while
all the others give small contribution to the binding energy.

\begin{table}[bth]
  \centering
  \begin{tabular}{cccc}
    \hline
    \hline
    class &  partial waves & $\ell_{\text{sum}}$ & $K_{iM}$\\
    \hline
    $C_1^L$ &  ${}^3S_1$                    & $\ell_{\text{sum}}=0$ & 14\\
    $C_2^L$ &  ${}^3D_1$,${}^5D_1$,${}^7D_1$& $\ell_5=2$,
    $\sum_{i=1,4}\ell_i=0$ & 12\\
    $C_3^L$ &  ${}^3S_1$                    & $\ell_{\text{sum}}=2$ & 10\\
    $C_4^L$ &  ${}^3D_1$,${}^5D_1$,${}^7D_1$& $\ell_{\text{sum}}=2$,
    not included in $C_2^L$ & 10\\
    $C_5^L$ &  ${}^1P_1$,${}^3P_1$,${}^5P_1$& $\ell_{\text{sum}}=2$ & 8\\
    $C_6^L$ &  ${}^5F_1$,${}^7F_1$,${}^7G_1$& $\ell_{\text{sum}}=4$ & 8\\
    \hline
    \hline
  \end{tabular}
  \caption{\label{tab:classli}
    Definition of the classes of hyperangular-spin-isospin states
    $\Psi^{KLSTJ\pi}_\alpha$ [see Eq.~(\ref{eq:hhst0})] used for
    the $\Li$ bound state as given in Ref.~\cite{Gnech2020}.
    The classes are defined by selecting particular values of the
    total orbital angular momentum $L$ and total spin $S$ (indicated
    by the spectroscopic notation
    $^{2S+1}L_J$),
    and the value of $\ell_{\text{sum}}=\ell_1+\cdots+\ell_5$,
    given in the second and third column, respectively.
    In the last column, the maximum $K$ value adopted in the
    expansion is reported for each class.}
\end{table}
In Tables~\ref{tab:classli} and~\ref{tab:classhe} we report the properties of
the HH states used to define a given class for, respectively, $\Li$ and $\Hes$.
For each class $i$ we also give the maximum value of $K$ we have adopted ($K_{iM}$).
A more detailed discussion of the class definition for $\Li$ can be found in Ref.~\cite{Gnech2020}.  
Here, we only note that in the case of $\Hes$ 
we divide the HH states in six different classes.
Classes $C_1^H$ and $C_2^H$ are the main components of
the $\Hes$ ground state, since they correspond to two-body correlated states having
$L\,$=$\,0$ and 2, respectively. For both of them we reach $K$ values up to $K_{1M}\,$=$\,K_{2M}\,$=$\,12$.
Classes $C_{3}^H$ and $C_{4}^H$ contain HH states that generate many-body
correlations for the $S$ and $D$ wave, respectively. For this reason, their
contribution to the binding energy is smaller and we stop at $K_{3M}\,$=$\,K_{4M}\,$=$\,10$.
Class $C_5^H$ contains HH states with $L\,$=$\,1$, which
are less important in the construction of the wave function.  We
therefore keep $K$ values up to $K_{5M}=8$ for these.
Finally, class $C_6^H$ consists of HH states with $L\,$=$\,3$. Their contribution
to the binding energy is tiny and so we select $K_{6M}\,$=$\,8$.

\begin{table}[bth]
  \centering
  \begin{tabular}{cccc}
    \hline
    \hline
    class &  partial waves & $\ell_{\text{sum}}$ & $K_{iM}$\\
    \hline
    $C_1^H$ &  ${}^1S_0$& $\ell_{\text{sum}}=0$ & 12\\
    $C_2^H$ &  ${}^5D_0$& $\ell_5=2$,\, $\sum_{i=1,4}\ell_i=0$ & 12\\
    $C_3^H$ &  ${}^1S_0$& $\ell_{\text{sum}}=2$ & 10\\
    $C_4^H$ &  ${}^5D_0$& $\ell_{\text{sum}}=2$,\, not included in $C_2^H$ & 10\\
    $C_5^H$ &  ${}^3P_0$& $\ell_{\text{sum}}=2$ & 8\\
    $C_6^H$ &  ${}^7F_0$& $\ell_{\text{sum}}=4$ & 8\\
    \hline
    \hline
  \end{tabular}
  \caption{\label{tab:classhe}
    Same as Table~\ref{tab:classli} but for $\Hes$.}
\end{table}

\section{Convergence of the HH expansion}\label{app:conv}

In this appendix we study the convergence of the $\Li$ and $\Hes$ binding
energies and discuss the extrapolation method.  The convergence is studied
class by class.  When studying the convergence of a generic class $C_i$, we
include in the expansion all the HH states with $K\leq K_i$ and then vary $K_i$
between a minimum value and $K_{iM}$.  For the other classes $C_j$ with
$j\ne i$, we include all HH states up to $K_{jM}$.
Note that for classes $C_1^L$ and $C_2^L$ ($C_1^H$ and $C_2^H$) in $\Li$ ($\Hes$),
because of the procedure used for the selection of the linearly independent
HH states, we cannot include, respectively, classes $C_3^L$ and $C_4^L$
($C_{3}^H$ and $C_{4}^H$).  The $\Li$ and $\Hes$ binding energies are
listed in Tables~\ref{tab:convli} and~\ref{tab:convhe}.
\begin{table}
  \centering
  \begin{tabular}{ccccccccccccccc}
    \hline
    \hline
    &  &  &  &  &  &\multicolumn{5}{c}{E-model} &\multicolumn{4}{c}{P-model} \\
    \cline{7-11}\cline{12-15}
    $K_1$ & $K_2$ & $K_{3}$ & $K_{4}$ & $K_5$ & $K_6$ & SRG$1.2$
    &SRG$1.5$ &SRG$1.8$ & SRG$2.0$ & bare &SRG$1.2$ &SRG$1.5$ &SRG$1.8$ & SRG$2.0$ \\
    \hline
    2  & 12 &    & 10 & 8 & 8 & 27.000 & 26.782 & 25.537 & 24.621 & 19.844 & 27.088 & 27.022 & 25.976 & 25.169\\  
    4  & 12 &    & 10 & 8 & 8 & 30.573 & 30.892 & 29.909 & 29.066 & 24.238 & 30.766 & 31.259 & 30.404 & 29.566\\
    6  & 12 &    & 10 & 8 & 8 & 31.645 & 32.468 & 31.845 & 31.152 & 26.619 & 31.857 & 32.872 & 32.365 & 31.632\\
    8  & 12 &    & 10 & 8 & 8 & 31.949 & 32.991 & 32.559 & 31.957 & 27.732 & 32.163 & 33.400 & 33.072 & 32.387\\
    10 & 12 &    & 10 & 8 & 8 & 32.057 & 33.185 & 32.822 & 32.254 & 28.279 & 32.271 & 33.594 & 33.331 & 32.662\\
    12 & 12 &    & 10 & 8 & 8 & 32.095 & 33.257 & 32.923 & 32.368 & 28.554 & 32.308 & 33.669 & 33.435 & 32.776\\
    14 & 12 &    & 10 & 8 & 8 & 32.108 & 33.284 & 32.960 & 32.410 & 28.725 & 32.322 & 33.696 & 33.474 & 32.819\\
    &    &    &    &   &   &        &        &        &        &        &        &        &        &       \\
    14 & 2  & 10 &    & 8 & 8 & 30.917 & 30.480 & 27.946 & 25.917 & 16.222 & 31.195 & 31.076 & 28.586 & 26.078\\
    14 & 4  & 10 &    & 8 & 8 & 31.555 & 31.643 & 29.593 & 27.800 & 18.602 & 31.808 & 32.191 & 30.199 & 27.957\\
    14 & 6  & 10 &    & 8 & 8 & 31.951 & 32.712 & 31.591 & 30.403 & 23.451 & 32.176 & 33.174 & 32.120 & 30.600\\
    14 & 8  & 10 &    & 8 & 8 & 32.038 & 33.038 & 32.373 & 31.542 & 26.315 & 32.256 & 33.469 & 32.881 & 31.832\\
    14 & 10 & 10 &    & 8 & 8 & 32.060 & 33.138 & 32.650 & 31.974 & 27.624 & 32.276 & 33.558 & 33.160 & 32.329\\
    14 & 12 & 10 &    & 8 & 8 & 32.068 & 33.177 & 32.765 & 32.160 & 28.265 & 32.283 & 33.593 & 33.277 & 32.554\\
    &    &    &    &   &   &        &        &        &        &        &        &        &        &       \\
    14 & 12 & 6  & 10 & 8 & 8 & 32.109 & 33.287 & 32.964 & 32.416 & 28.734 & 32.323 & 33.698 & 33.476 & 32.823\\
    14 & 12 & 8  & 10 & 8 & 8 & 32.142 & 33.333 & 33.016 & 32.469 & 28.779 & 32.355 & 33.744 & 33.528 & 32.873\\
    14 & 12 & 10 & 10 & 8 & 8 & 32.158 & 33.358 & 33.047 & 32.501 & 28.813 & 32.371 & 33.769 & 33.558 & 32.904\\
    &    &    &    &   &   &        &        &        &        &        &        &        &        &       \\
    14 & 12 & 10 & 4  & 8 & 8 & 32.074 & 33.187 & 32.779 & 32.175 & 28.285 & 32.290 & 33.604 & 33.293 & 32.572\\
    14 & 12 & 10 & 6  & 8 & 8 & 32.124 & 33.276 & 32.904 & 32.319 & 28.480 & 32.338 & 33.691 & 33.418 & 32.720\\
    14 & 12 & 10 & 8  & 8 & 8 & 32.150 & 33.336 & 33.003 & 32.442 & 28.683 & 32.364 & 33.748 & 33.516 & 32.843\\
    14 & 12 & 10 & 10 & 8 & 8 & 32.158 & 33.358 & 33.047 & 32.501 & 28.813 & 32.371 & 33.769 & 33.558 & 32.904\\
    &    &    &    &   &   &        &        &        &        &        &        &        &        &       \\
    14 & 12 & 10 & 10 & 2 & 8 & 32.078 & 33.181 & 32.769 & 32.163 & 28.228 & 32.303 & 33.611 & 33.292 & 32.567\\
    14 & 12 & 10 & 10 & 4 & 8 & 32.132 & 33.287 & 32.917 & 32.332 & 28.462 & 32.348 & 33.703 & 33.418 & 32.727\\
    14 & 12 & 10 & 10 & 6 & 8 & 32.151 & 33.335 & 33.000 & 32.437 & 28.656 & 32.364 & 33.748 & 33.512 & 32.837\\
    14 & 12 & 10 & 10 & 8 & 8 & 32.158 & 33.358 & 33.047 & 32.501 & 28.813 & 32.371 & 33.769 & 33.558 & 32.904\\
    &    &    &    &   &   &        &        &        &        &        &        &        &        &       \\
    14 & 12 & 10 & 10 & 8 & 4 & 32.145 & 33.314 & 32.953 & 32.371 & 28.530 & 32.359 & 33.730 & 33.470 & 32.773\\
    14 & 12 & 10 & 10 & 8 & 6 & 32.154 & 33.342 & 33.007 & 32.442 & 28.662 & 32.367 & 33.755 & 33.521 & 32.844\\
    14 & 12 & 10 & 10 & 8 & 8 & 32.158 & 33.358 & 33.047 & 32.501 & 28.813 & 32.371 & 33.769 & 33.558 & 32.904\\
    \hline
    \hline
  \end{tabular}
  \caption{\label{tab:convli} Convergence of the $\Li$ binding energy
    for the different classes $C_1^L$--$C_6^L$, into which the HH
    states have been divided. All results are in MeV units.}
\end{table}
\begin{table}
  \centering
  \begin{tabular}{ccccccccccccccc}
    \hline
    \hline
    &  &  &  &  &  &\multicolumn{5}{c}{E-model} &\multicolumn{4}{c}{P-model} \\
    \cline{7-11}\cline{12-15}
    $K_1$ & $K_2$ & $K_{3}$ & $K_{4}$ & $K_5$ & $K_6$ & SRG$1.2$
    &SRG$1.5$ &SRG$1.8$ & SRG$2.0$ & bare & SRG$1.2$ &SRG$1.5$ &SRG$1.8$ & SRG$2.0$ \\                                               
    \hline
    2  & 12 &    & 10  & 10 & 6 & 24.114 & 24.294 & 23.330 & 22.500 & 17.961 & 23.822 & 23.905 & 22.953 & 22.117\\        
    4  & 12 &    & 10  & 10 & 6 & 27.176 & 27.585 & 26.749 & 25.985 & 21.585 & 27.253 & 27.742 & 26.969 & 26.176\\        
    6  & 12 &    & 10  & 10 & 6 & 28.295 & 29.194 & 28.713 & 28.108 & 24.054 & 28.419 & 29.449 & 29.053 & 28.398\\        
    8  & 12 &    & 10  & 10 & 6 & 28.651 & 29.764 & 29.457 & 28.933 & 25.149 & 28.789 & 30.046 & 29.819 & 29.209\\        
    10 & 12 &    & 10  & 10 & 6 & 28.802 & 30.010 & 29.776 & 29.285 & 25.721 & 28.943 & 30.301 & 30.143 & 29.551\\        
    12 & 12 &    & 10  & 10 & 6 & 28.870 & 30.123 & 29.921 & 29.444 & 26.024 & 29.013 & 30.419 & 30.295 & 29.712\\        
    &    &    &     &    &   &        &        &        &        &        &        &        &        &       \\
    12 & 4  & 10 &     & 10 & 6 & 28.442 & 28.847 & 27.274 & 25.796 & 17.845 & 28.609 & 29.233 & 27.696 & 25.831\\        
    12 & 6  & 10 &     & 10 & 6 & 28.720 & 29.586 & 28.652 & 27.593 & 21.218 & 28.871 & 29.917 & 29.019 & 27.647\\        
    12 & 8  & 10 &     & 10 & 6 & 28.805 & 29.900 & 29.383 & 28.647 & 23.807 & 28.950 & 30.205 & 29.739 & 28.786\\        
    12 & 10 & 10 &     & 10 & 6 & 28.833 & 30.011 & 29.669 & 29.080 & 25.048 & 28.976 & 30.308 & 30.028 & 29.283\\        
    12 & 12 & 10 &     & 10 & 6 & 28.844 & 30.058 & 29.799 & 29.284 & 25.692 & 28.986 & 30.352 & 30.163 & 29.528\\        
    &    &    &     &    &   &        &        &        &        &        &        &        &        &       \\
    12 & 12 & 4  & 4   & 10 & 6 & 28.843 & 30.058 & 29.802 & 29.288 & 25.705 & 28.986 & 30.355 & 30.172 & 29.541\\
    12 & 12 & 6  & 6   & 10 & 6 & 28.860 & 30.093 & 29.857 & 29.357 & 25.833 & 29.003 & 30.389 & 30.230 & 29.616\\
    12 & 12 & 8  & 8   & 10 & 6 & 28.878 & 30.132 & 29.923 & 29.440 & 25.982 & 29.021 & 30.426 & 30.296 & 29.703\\
    12 & 12 & 10 & 10  & 10 & 6 & 28.887 & 30.151 & 29.956 & 29.483 & 26.074 & 29.029 & 30.445 & 30.329 & 29.749\\
    &    &    &     &    &   &        &        &        &        &        &        &        &        &       \\
    12 & 12 & 10 & 10  & 2  & 6 & 28.719 & 29.831 & 29.501 & 28.956 & 25.251 & 28.884 & 30.153 & 29.891 & 29.226\\
    12 & 12 & 10 & 10  & 4  & 6 & 28.799 & 29.961 & 29.662 & 29.127 & 25.438 & 28.952 & 30.270 & 30.042 & 29.388\\
    12 & 12 & 10 & 10  & 6  & 6 & 28.856 & 30.081 & 29.842 & 29.340 & 25.766 & 29.002 & 30.380 & 30.217 & 29.601\\
    12 & 12 & 10 & 10  & 8  & 6 & 28.876 & 30.127 & 29.916 & 29.430 & 25.939 & 29.020 & 30.423 & 30.288 & 29.693\\
    12 & 12 & 10 & 10  & 10 & 6 & 28.887 & 30.151 & 29.956 & 29.483 & 26.074 & 29.029 & 30.445 & 30.329 & 29.749\\
    &    &    &     &    &   &        &        &        &        &        &        &        &        &       \\
    12 & 12 & 10 & 10  & 10 & 4 & 28.886 & 30.148 & 29.948 & 29.472 & 26.044 & 29.029 & 30.443 & 30.321 & 29.736\\
    12 & 12 & 10 & 10  & 10 & 6 & 28.887 & 30.151 & 29.956 & 29.483 & 26.074 & 29.029 & 30.445 & 30.329 & 29.749\\
    \hline
    \hline
  \end{tabular}
  \caption{\label{tab:convhe} Same as Table~\ref{tab:convli} but for
    $\Hes$.}
\end{table}

We assume that for each class of convergence the behavior of
the binding energy as function of $K$ is exponential, namely
\begin{equation}\label{eq:fitb}
  B_i(K)=B_i(\infty)+a_i\,{\rm e}^{-b_iK}\,,
\end{equation}
where $B_i(\infty)$ is the asymptotic binding energy of class
$C_i$ as $K\longrightarrow\infty$.
The parameters $a_i$ and $b_i$ depend on the interaction model and on
the specific class of HH states we are studying. The values
of $B_i(K)$ are those reported in Tables~\ref{tab:convli}
and~\ref{tab:convhe}. By defining the function
\begin{equation}
  \Delta_i(K)=B_i(K)-B_i(K-2)\,,
\end{equation}
it is possible to compute, for each class, the ``missing'' binding energy due to the
truncation of the expansion to a finite $K_{iM}$ as illustrated in Ref.~\cite{Viviani2005},
namely,
\begin{equation}\label{eq:dbi}
  \left(\Delta B\right)_i=\sum_{K= K_{iM}+2,K_{iM}+4,\dots}\Delta_i(K)\,.
\end{equation}
By using Eq.~(\ref{eq:fitb}), we obtain
\begin{equation}
  \left(\Delta B\right)_i=\Delta_i(K_{iM})\frac{1}{{\rm e}^{2b_i}-1}\,.
\end{equation}
The ``total missing'' binding energy is then computed as
\begin{equation}\label{eq:dbt}
  \left(\Delta B\right)_T=\sum_{i=1,6}\Delta_i(K_{iM})
  \frac{1}{{\rm e}^{2b_i}-1}\,.
\end{equation}
In order to determine the coefficients $b_i$ for each class,
we proceed as follows. 
For classes $C_1^L$, $C_2^L$, and $C_1^H$, we estimate the $b_i$
by performing a fit to the binding energy values of Tables~\ref{tab:convli} and~\ref{tab:convhe},
using Eq.~(\ref{eq:fitb}).  We propagate the error on the resulting $b_i$
to compute the error on $\left(\Delta B\right)_i$.  For classes $C_5^L$,
$C_2^H$, and $C_5^H$, the quality of the fit is not good enough to obtain
a sensible estimate. In these cases, we consider a reasonable range for $b_i$,
\begin{equation}
  \min{\{b_i^0,b_i^1\}}\leq b_i \leq \max{\{b_i^0,b_i^1\}}\ ,
\end{equation}
where $b_i^0$ and $b_i^1$ are computed from
\begin{equation}
  \frac{\Delta_i(K_{iM}-2)}{\Delta_i(K_{iM})}={\rm e}^{2b_i^0}\,,\qquad
  \frac{\Delta_i(K_{iM}-4)}{\Delta_i(K_{iM}-2)}={\rm e}^{2b_i^1}\ .
\end{equation}
We use the central value of the interval as the best estimate, and the
range as error bar.  For classes $C_3^L$, $C_4^L$, and $C_3^H+C_4^H$, we
simply estimate $b_i$ from
\begin{equation}
  \frac{\Delta_i(K_{iM}-2)}{\Delta_i(K_{iM})}={\rm e}^{2b_i}\,.
\end{equation}
In such cases, we use these $b_i$ to obtain the missing energy, and
estimate the error as half of this missing energy.
Finally, for classes $C_6^L$ and $C_6^H$ it is not possible to
obtain reliable values for the $b_i$. Therefore, we estimate the
missing binding energy as $\Delta_i(K_{iM})$ and the error as
half of it.  In Tables~\ref{tab:mbeli} and~\ref{tab:mbehe} we report the missing
binding energy with the associated error
for each of the six classes we have considered.
\begin{table}
  \centering
  \begin{tabular}{cccccccccc}
    \hline
    \hline
    &\multicolumn{5}{c}{E-model}&\multicolumn{4}{c}{P-model} \\
    \cline{2-6}\cline{7-10}
    & SRG$1.2$ &SRG$1.5$ &SRG$1.8$ & SRG$2.0$ & bare & SRG$1.2$ &SRG$1.5$ &SRG$1.8$ & SRG$2.0$ \\
    \hline
    $C_1^L$ & 0.007(0) & 0.016(0) & 0.022(0) & 0.025(0) & 0.175(0) & 0.008(0) & 0.016(0) & 0.023(1) & 0.026(1) \\
    $C_2^L$ & 0.003(0) & 0.018(2) & 0.066(5) & 0.118(7) & 0.557(19)& 0.002(0) & 0.016(2) & 0.071(4) & 0.173(10)\\
    $C_3^L$ & 0.015(8) & 0.030(15)& 0.046(23)& 0.049(24)& 0.105(53)& 0.016(8) & 0.030(15)& 0.041(20)& 0.051(25)\\
    $C_4^L$ & 0.004(2) & 0.013(6) & 0.035(18)& 0.054(27)& 0.231(116)&0.003(1) & 0.012(6) & 0.032(16)& 0.060(30)\\
    $C_5^L$ & 0.004(0) & 0.020(1) & 0.061(1) & 0.102(3) & 0.319(25) &0.005(1) & 0.019(1) & 0.074(60)& 0.123(24)\\
    $C_6^L$ & 0.008(4) & 0.032(16)& 0.080(40)& 0.118(59)& 0.302(151)&0.008(4) & 0.030(15)& 0.074(37)& 0.120(60)\\
    \hline
    Tot.    & 0.033(9) & 0.113(23)& 0.288(50)& 0.442(70)& 1.515(200)&0.034(9) & 0.107(22)& 0.292(75)& 0.527(76)\\
    \hline
    \hline
  \end{tabular}
  \caption{\label{tab:mbeli} Missing binding energies corresponding to the
  E and P interaction models, obtained for each of the
  six classes of convergence we have considered for $\Li$.  In parentheses,
    we report the errors on the extrapolation; note that $(0)$ indicates
    that the error does not affect the last digit reported in the result.}
\end{table}
\begin{table}
  \centering
  \begin{tabular}{cccccccccc}
    \hline
    \hline
    &\multicolumn{5}{c}{E-model}&\multicolumn{4}{c}{P-model} \\
    \cline{2-6}\cline{7-10}
    & SRG$1.2$ &SRG$1.5$ &SRG$1.8$ & SRG$2.0$ & bare  SRG$1.2$ &SRG$1.5$ &SRG$1.8$ & SRG$2.0$  \\
    \hline
    $C_1^H$ & 0.051(1) & 0.088(3) & 0.111(3) & 0.121(3) & 0.333(4) & 0.052(2) & 0.091(4) & 0.116(6) & 0.123(7)\\
    $C_2^H$ & 0.006(1) & 0.030(5) & 0.095(13)& 0.161(21)& 0.643(52)& 0.006(1) & 0.028(4) & 0.104(15)& 0.213(26)\\
    $C_3^H$ & \multirow{2}{*}{0.009(4)} & \multirow{2}{*}{0.018(9)}& \multirow{2}{*}{0.033(16)}& \multirow{2}{*}{0.046(23)}
    & \multirow{2}{*}{0.148(74)} & \multirow{2}{*}{0.006(3)} & \multirow{2}{*}{0.020(10)}&\multirow{2}{*}{0.033(17)}&
    \multirow{2}{*}{0.052(26)}\\
    $C_4^H$ &  &  & & & & & & &\\
    $C_5^H$ & 0.009(4) & 0.020(6) & 0.036(11)& 0.054(22)& 0.251(213) & 0.007(2) & 0.018(5) & 0.040(16) & 0.061(26) \\
    $C_6^H$ & 0.002(1) & 0.006(3) & 0.016(8) & 0.022(11)& 0.060(30) & 0.002(1)  & 0.004(2) & 0.016(8) & 0.026(13)\\
    \hline
    Tot.    & 0.078(7) & 0.162(13)& 0.292(25)& 0.405(40)& 1.434(233) & 0.072(5) & 0.161(13)& 0.309(29) & 0.474(47)\\
    \hline
    \hline
  \end{tabular}
  \caption{\label{tab:mbehe} Same as Table~\ref{tab:mbeli} but for $\Hes$.}
\end{table}
\bibliography{bibliography}

\end{document}